\documentclass[prb,reprint,aps,floatfix,showpacs,superscriptaddress,longbibliography]{revtex4-1}
\usepackage{amssymb}
\usepackage{bbm}
\usepackage{dsfont}

\usepackage{amsmath}
\usepackage{graphicx}
\usepackage[caption=false]{subfig}
\usepackage[colorlinks=true,linkcolor=blue,anchorcolor=red,citecolor=blue,urlcolor=blue]{hyperref}
\begin{document}
\title{
%Anomalous temperature dependence of 
Ballistic magnetotransport in graphene}
\author{Ke Wang}
\email{kewang@umass.edu}
\affiliation{%
	Department of Physics, University of Massachusetts, Amherst, MA 01003, USA}
%\author{M. E. Raikh}
%\affiliation{Department of Physics and Astronomy, University of Utah, Salt Lake City, UT 84112, USA}
\author{T. A. Sedrakyan}%
\email{tsedrakyan@umass.edu}
\affiliation{%
	Department of Physics, University of Massachusetts, Amherst, MA 01003, USA}

\date{\today}
\begin{abstract} 
We report that a perpendicular magnetic field introduces an anomalous interaction correction, $\delta \sigma$, to the static conductivity of doped graphene in the ballistic regime. The correction implies that the magnetoresistance, $\delta \rho_{xx}$  scales inversely with temperature  $\delta \rho_{xx}(T) \propto 1/T$ in a parametrically large interval. {  When the disorder is scalar-like}, the $\propto 1/T$ behavior is the leading contribution in the crossover between diffusive regime exhibiting weak localization and quantum magnetooscillations. The behavior originates from the field-induced breaking of the chiral symmetry of Dirac electrons around a single valley. The result is specific for generic two-dimensional Dirac materials which deviate from the half-filling. We conclude by proposing magnetotransport experiments, which have the capacity to detect the nature of impurities and defects in high-mobility Dirac monolayers such as recently fabricated ballistic graphene samples. 
%We also discuss the observable effects in magnetotransport experiments on high-mobility doped graphene monolayers. 
\end{abstract}
\maketitle
{\em{Introduction}.}
Two well-established regimes characterize low-temperature magnetoresistance in a two-dimensional metallic system: Weak localization\cite{potp80Hikamai,prb80Altshuler} (WL) and Shubnikov–de Haas (SdH) oscillations. WL dominates in the low field limit $\omega_0 \tau < (k_F L)^{-1}$. Here $\omega_0$ is the cyclotron-frequency, $\tau$ is the impurity scattering time, and $L=v_F \tau$ is the mean free path. This regime is reached when the magnetic flux threading the area $L^2/2$ is smaller than the flux quantum\cite{ssc94Dyakonov,france94Cassam}. In the high field limit $\omega_0 \tau > 1$, the spectrum is fully quantized into the Landau levels, and SdH oscillations become the dominating effect. The crossover between two limits is $(k_F L)^{-1}<\omega_0 \tau<1$, where the magnetic field B is non-quantizing. In this regime, the electron-electron interaction (EEI) are believed to play a significant role\cite{prl08crossover}. Namely, the interaction correction to the conductivity induce the B-dependence in the resistivity via the relation,
\begin{eqnarray}
\label{relation}
\delta \rho_{\it{int}}\simeq \rho_0^2 (\omega_0^2 \tau^2 -1)\delta \sigma_{\it{int}}.
\end{eqnarray}
Here $\rho_0$ is the Drude resistivity and $\delta \sigma_{int}$ is the interaction correction to the longitudinal conductivity.%, known as $\sim T\tau$ in the ballistic limit $T\tau >1$ and $\sim \ln(T\tau)$ in the diffusive limit $T\tau < 1$.  Behaviors in both limits originate from coherent scatterings of Friedel oscillations (FO) \cite{52Friedel}.%, the behavior of the electron density in the large distance $r\gg k^{-1}_F$ from an impurity.
%Here $\delta \sigma_{\it{int}}$ originates from coherent scatterings of Friedel oscillations \cite{52Friedel}, the behavior of the electron density in the large distance $r\gg k^{-1}_F$ from an impurity. The temperature dependence\cite{prb86Gold,01prbzala} in $\delta \sigma_{int}$ is $\sim T\tau$ in the ballistic limit $T\tau >1$ and $\sim \ln(T\tau)$ in the diffusive limit $T\tau < 1$. The coherent scattering is illustrated in Fig.~\ref{scattering}.

%The 2DEG and highly-doped graphene are both metallic system. One may naively exepect same behavior for 2DEG and hdg. However, it is not true. Experimetally, ...Theoretically,  we show that  non-quantizing magnetic field leads to both non-trivial ....Main reason is that ... break cs symmetry.In this letter, we established

At the non-quantizing regime, magnetoresistance in the doped graphene has been widely studied in experiments\cite{prb10Kozikov,prb11Jouault,prl12Jobster,prb14Jabakhanji,nc15Gopinadhan} in the last decade while theoretical investigations are still absent. One may expect that the B-dependence in $\delta \rho_{\it{int}}$ is simply product of $\rho^2\omega_0^2 \tau^2 $ and zero field performance in $\delta \sigma_{\it{int}}$. However, it is not the complete story. We have shown that the non-quantizing field on Dirac electrons has non-trivial effects on FO\cite{prb20wang} and many-body physics\cite{wang2021}.% indicate that   %Namely, $\delta \sigma_{\it{int}}$ itself can contain field-dependent corrections,

In this letter, we report that $\delta \sigma_{\it{int}}$ itself can carry field-dependent corrections and thus leads to non-trivial magnetoresistance in graphene. In the ballistic regime $T\tau>1$ of the doped graphene\cite{nl16Banszerus,nc15Gopinadhan}, we  find  
\begin{eqnarray} 
\label{int2}
\delta \sigma_{\it{int}}\simeq   \lambda_0\frac{e^2\tau}{\pi  } \Big( {t} T- {p} \frac{\omega_0^2}{48T}  \Big).
\end{eqnarray}
Here $\lambda_0$ is the dimensionless interaction parameter, and ${t}$, ${p}$ are dimensionless parameters determined by the disorder potential. 
Information about ${t}$ and ${p}$ can be extracted from the zero-bias anomaly\cite{prb07Glazman,wang2021} of tunneling density of states. 
The correction is present in a wide parameter range, where $\max( \omega_0, \tau^{-1}) <T<E_F$. Here $E_F$ is the Fermi energy. From Eq.~\ref{relation}, the field-dependent correction to the resistivity reads\cite{footnote3},
\begin{eqnarray}
\label{resistivity}
\delta \rho_{\it{int}}(B)-\delta \rho_{\it{int}}(0) \simeq \lambda_0 \omega_0^2  \frac{ e^2\tau^2 \rho^2_0}{\pi  }   \left({t} T  \tau +  \frac{{p} }{48T\tau }\right).\nonumber\\
\end{eqnarray}
Temperature dependence of magnetoresistance in  Eq.~\ref{resistivity} highly depends on the ratio, $p/t$, of two disorder parameters that will be defined below. Generally, the ratio $p/t$ can be any real number larger than $-1/2$. One prominent case is the scalar-like disorder potential, for which ${p}/{t}$ is $\rightarrow +\infty$. To ensure the second term is not always subleading, we will focus on $p/t> 1$, where the disorder can be regarded as a perturbation around a scalar-like potential. When $1<T\tau< \sqrt{p/t}$, the temperature dependence in magnetoresistance becomes reciprocal instead of being linear. Therefore, the parabolic curve in magnetoresistance becomes more flattened when $T$ increases. See Fig.~\ref{resisitivity_plot}.
Below, we present a qualitative explanation of the observed effect. 

  \begin{figure}
	\centering
	\includegraphics[scale=0.55]{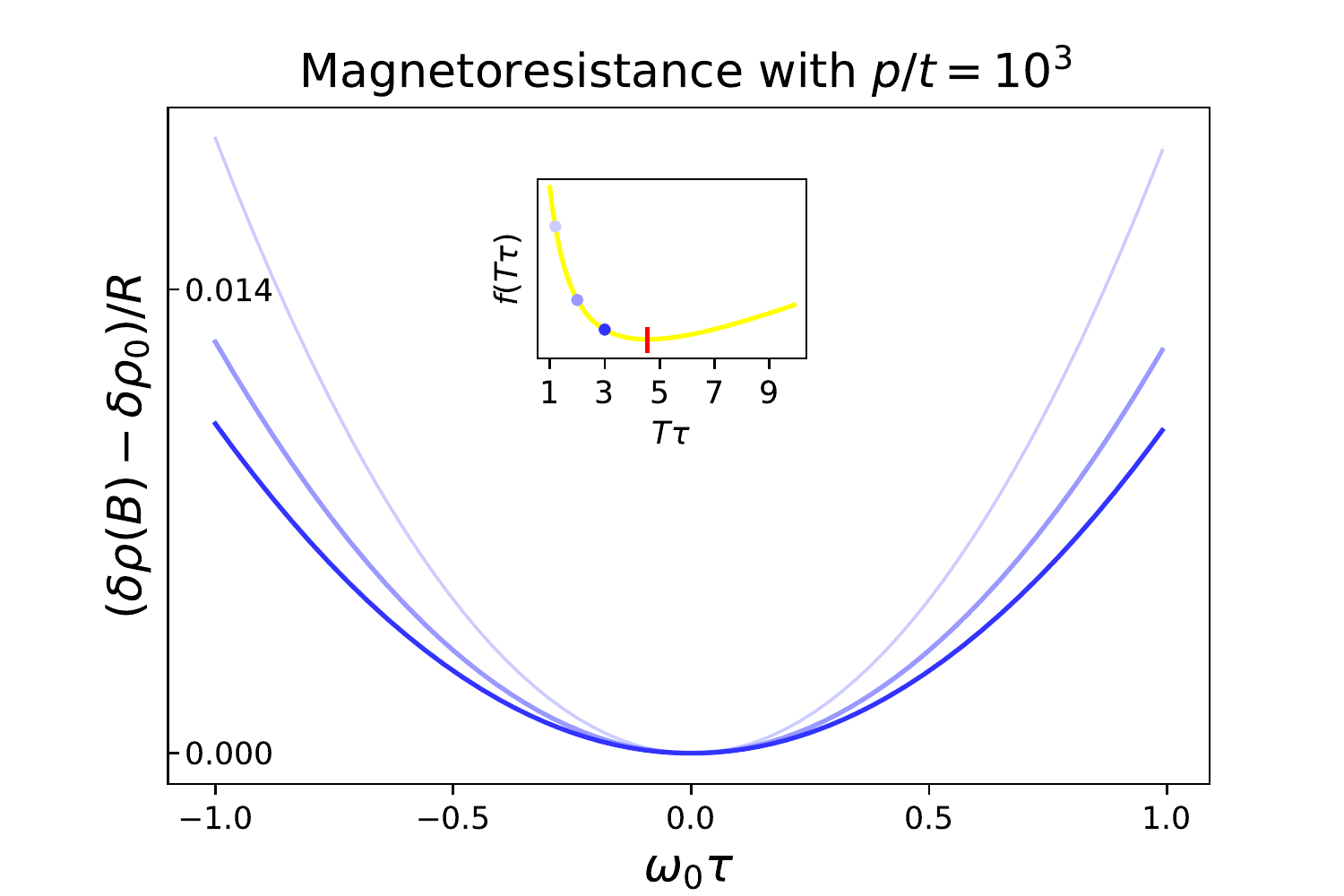} 
	\caption{Magnetoresistance, $[\delta \rho_{\it{int}}(B)-\delta \rho_{\it{int}}(0)  ]/R$, is plotted versus the dimensionless variable $\omega_0\tau$. The sign of $\omega_0\tau$ indicates the direction of the magnetic field and $R\equiv \lambda_0     { e^2  \rho^2_0}/{\pi  }$. Each curve correspond to the resistance plotted at a corresponding temperature shown by (blue) dots in the inset. From light to dark curves, the temperature is increasing while the curvature is decreasing. Values of temperature are pointed out in the inset. The inset depicts the function $f(T\tau)= {t} T  \tau +  {{p} }/{48T\tau }$ from Eq.~\ref{resistivity}.  The (red) vertical bar locates the minimum of the function. }
	\label{resisitivity_plot}
\end{figure} 
%Weak localization\cite{potp80Hikamai,prb80Altshuler,prl02Suzuura,prl06Khveshchenko,prl06Mccann,prl06Morozov}

%  renormalizes the transport relaxation time, $\tau$. Concretely, coherent scatterings of Friedel oscillations leads to non-trivial temperature dependence in the longitudnal and static conductivity, $\delta \sigma_{xx}$. Namely, the temperature dependenc\cite{prb86Gold,01prbzala} is $\sim T\tau$ in the ballistic limit ($T\tau \gg 1$) and $\sim \ln(T\tau)$ in the diffusive limit  ($T\tau \ll 1$). The coherent scattering is illustrated in Fig.~\ref{scattering}. 
  
 {\it Qualitative discussion.} 
 Coherent scatterings off Friedel oscillations of electron density at distances $r\gg k^{-1}_F$ from an impurity renormalize the transport relaxation time.
 The coherent scattering is illustrated in Fig.~\ref{scattering}. This process, leads to non-trivial temperature dependence\cite{prb86Gold,01prbzala} in $\delta \sigma_{int}$. In two-dimensional electron gas (2DEG), $\delta \sigma_{int}$ is  $\sim T\tau$ in the ballistic limit $T\tau >1$ and $\sim \ln(T\tau)$ in the diffusive limit $T\tau < 1$.

 Dirac nature of electrons in graphene \cite{Tan2007,08sscKim,prl07,prb07das,prl08Kim,rmp09Castroneto,rmp11Dassarma,rmp11Goerbig,rmp12Kotov,prb19Saurabh,nature19,prl20Pan,prb21Narozhny,prb20Wagner,prb21Mikhailov,prb21Zhu,prb21Kamada,prl20Real,prb21Sharma} can enrich the process of coherent scatterings because of the Berry phase $\pi$ and chiral symmetry of Dirac electrons. Note that backscatterings
 off a single impurity can be classified into two types of Feynman diagrams. The first one is a loop type, giving Friedel oscillations. See inset (a) in Fig.~\ref{scattering}. Here, the Berry phase $\pi$ of Dirac electron leads to a faster decaying FO\cite{prl06Cheianov}. The second one is a vertex type diagram, yielding the correction to the density matrix. See inset (b) of Fig.~\ref{scattering}. Here, the matrix structure of Dirac electron induces sensitivity of the vertex correction to the nature of disorder\cite{prb07Glazman}. Two properties together lead to the well-known result that the temperature dependence in the conductivity in the ballistic limit is still $\sim T\tau$ but very sensitive to the disorder\cite{prl06Cheianov}. Importantly, if the disorder is scalar-like, the leading temperature behavior $\sim T\tau$ vanishes.

  The presence of a weak magnetic field changes the scenario for both backscatterings in (a) and (b) from the inset of Fig.~\ref{scattering}. The persistent FO emerges from loop correction\cite{prb20wang}, 
  \begin{eqnarray}\label{FO}\delta n(r)=\frac{g k_F  }{2\pi^2v_F r^2}\Biggl[\frac{1}{k_F r}\cos\Bigl(2k_Fr-\frac{ r^3}{12k_F l^4}\Bigr)\nonumber\\+2\varphi^2(r)\sin\Bigl(2k_Fr-\frac{  r^3}{12k_F l^4}\Bigr)\Biggr].\end{eqnarray}
Here the parameter $g$ is defined in terms of the impurity potential, $\hat V_{\mathbf{r}} $, as $g=\text{tr}\int d^2r \hat V_{\mathbf{r}}/4$, $\varphi(r)= \omega_0 r/2v_F$ and $l$ is the magnetic length. The $\varphi(r)$ is the half of the angle of the arc, corresponding to the Dirac electron traveling from $\mathbf{0}$ to $\mathbf{r}$ in a weak magnetic field. See Fig.~\ref{scattering}. The value $\varphi(r)$ reflects the strength of chiral symmetry breaking semi-classically\cite{IOP12Gordon,prb20wang}. A similar correction also emerges for the vertex correction\cite{wang2021}. To evaluate the effect of the magnetic field on other physical processes\cite{prb10Raman} for Dirac electrons, employing the chiral-symmetry breaking phase $\varphi(r)$ could be essential as it could lead to novel observable effects.

As the next step,
we will consider the transport relaxation time and see that the incorporation of $\varphi^2(r)$ into the estimate of the relaxation time can generate the correction in Eq.~\ref{int2}. It will help us to qualitatively extract the temperature behavior of magnetoconductivity from its relation to the transport time\cite{mahan}.

At first, let us estimate the relaxation time at zero-field, where a linear temperature dependence emerges:
 \begin{eqnarray} 
 \label{1}
 \frac{1}{\tau}=\int \frac{d\theta}{2\pi} (1-\cos \theta) |f_0+  f_1(\theta)|^2.
 \end{eqnarray}
 Here $f_0$ and $f_1$ are respectively the scattering amplitudes off impurities and impurity-induced potentials. In the absence of the magnetic field, according to Refs.~\onlinecite{01prbzala} and ~\onlinecite{prl08Raikh}, the function $f_1$ can be cast as an integral $f_1(\theta)=\int dr F(r)$ and
 \begin{eqnarray}
 \label{2}
 F(r) =- {\lambda_0 g}  \int_0^{+\infty} dr \frac{r_T}{\sinh r/r_T} \sin (2k_Fr) J_0 ({q}{r}).
  \end{eqnarray} 
Here $r_T=  v_F/(2\pi T)$ is the thermal length, $|q|=2k_F \sin \theta/2$ and $J_0$ is the zero Bessel function. The coefficient $\lambda_0$ is the dimensionless interaction parameter and the main contribution to Eq.~\ref{1} comes from the region $\theta \sim \pi$. One can expand $\theta=\pi+\delta \theta$ and $q\simeq 2k_F-k_F \delta \theta^2$. The condition $k_F\delta \theta ^2 r_T\sim 1$ translates into $\delta \theta\sim (k_F r_T)^{-1/2}$.
With the asymptotic expression of Bessel function, the power counting in the integral becomes $r^{-3/2}$ when $r<r_T$.  When $\delta \theta<(k_F r_T)^{-1/2}$, the integral in Eq.~\ref{2}, gives $\sim (k_F r_T)^{-1/2}$. Thus the integral in Eq.~\ref{1} is estimated by $(k_F r_T)^{-1}$. This indicates that the interaction correction to $\tau$ is proportional to $T$ and  the corresponding correction to $\delta \sigma_{int}$ is also linear in $T$.

  \begin{figure}
	\centering
	\includegraphics[scale=0.235]{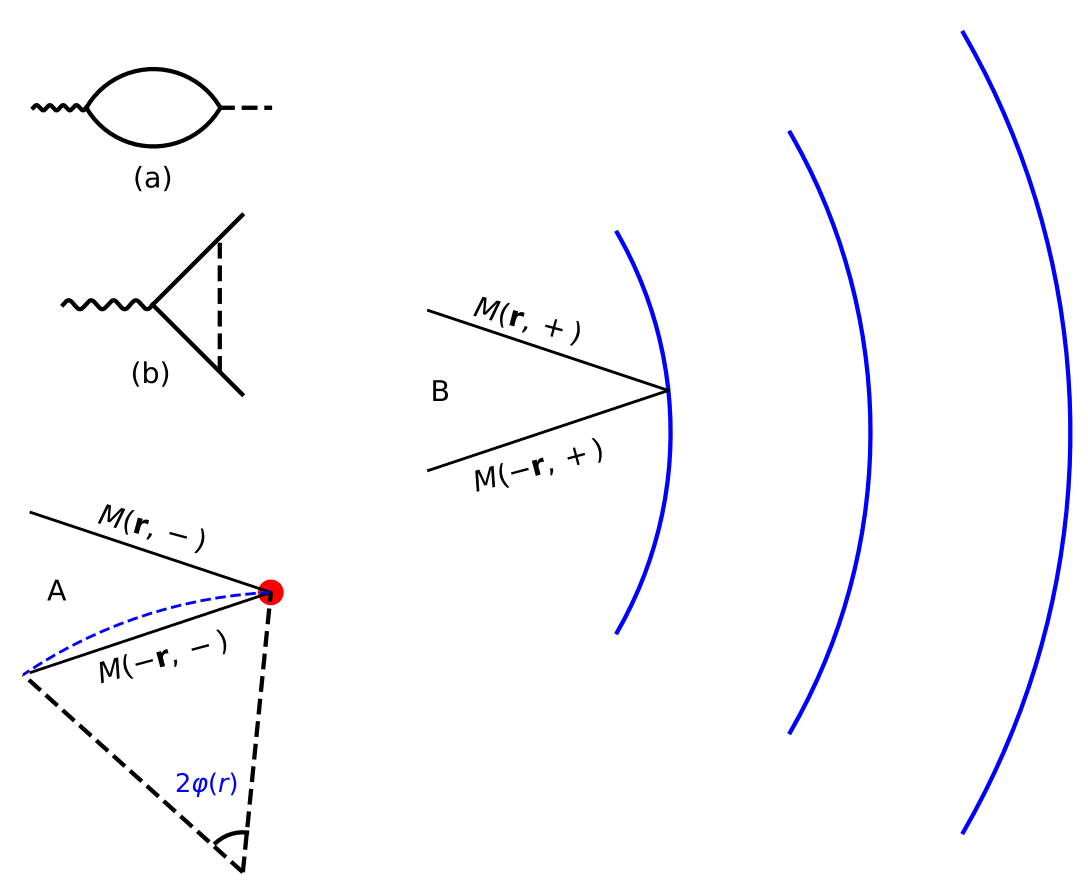} 
	\caption{Coherent scatterings between A and B paths: A is the path of backscattering off an impurity while B is the path when electrons hit the Friedel oscillations (or modulation of density matrix introduced by impurities), presented by blue curves. In the presence of a magnetic field, the path is curved, shown by the dashed arc. The angle of the arc is $2\varphi(r)=\omega_0 r/v_F$. Due to the Dirac nature of electrons, each propagator carries a matrix $M$.
	The inset plots two types of backscatterings off an impurity : (a) the loop type that creates Friedel oscillations. (b) the vertex type that creates correction to the density matrix.  }
	\label{scattering}
\end{figure} 
 %We start from the discussion that a linear  temperature dependence can be observed in the coherent scatterings' correction to the relaxation time,

In the presence of a magnetic field, the trajectories of electrons are curved, and the chiral symmetry of Dirac electrons is broken. Thus the suppressed backscattering is enhanced by the magnetic field.  The incorporation of the symmetry-breaking effect leads to field-dependent correction to the scattering amplitude. Namely, $f_1 \rightarrow f_1+\delta f_1$ and $\delta f_1$ is given by $\int dr F(r) \varphi^2(r) $. Here $\varphi^2(r)$ changes power counting to $r^{1/2}$ and the integral gives $\sim\omega_0^2 (k_F r_T)^{3/2}$. The $\theta$-integral remains the same. Thus the field-dependent corrections to $\tau$ and $\delta \sigma_{int}$ are proportional to $\omega_0^2 T^{-1}$. 
  
Below, we rigorously trace the current-current correlation function to derive the temperature-dependence in $\delta \sigma_{int}$.  
 %When Dirac electron travels along the curved-trajectory, the angle of the arc, $\varphi(r)$, reflects the strength of chiral-symmetry breaking (footnote here). See Fig.~ and note $\varphi(r)= \omega_0 r/v_F$.
 
 %The temperature dependence, in the conductivity originating from Friedel oscillations, is accumulated at the large distance $r\gg k^{-1}_F $. It was shown that the scale for 2D electron's $T\tau$-behavior is around $v_F/T$.
  
  {\it Magnetoconductivity from Kubo formula. }
  The static conductivity can be evaluated from the current-current correlation function\cite{mahan}. Namely, 
  $\sigma_{\alpha,\beta}
  =\lim_{\omega\rightarrow 0}\frac{i}{\omega} \Pi_{\alpha,\beta}(\omega)$. Here the $\Pi_{\alpha,\beta}(\omega)$ is obtained by taking analytic continuation of current-current correlation function $\Pi_{\alpha,\beta}(i\Omega_n)$ via $i\Omega_n \rightarrow \omega$, $
  \Pi_{\alpha,\beta}(i\Omega_n)= \int_0^{1/T} d\tau  \langle T_\tau \hat{j}_\alpha(\tau) \hat{j}_\beta(0)  \rangle e^{i\Omega_n\tau} $
  Here $\hat{j}_\alpha(\tau)$, $\alpha=1,2$, is the current operator at imaginary time $\tau$, $\omega_n=2\pi T n$ is the bosonic Matsubara frequency and $i\Omega_n \rightarrow \omega$ represents the analytic continuation. 
  
  At finite doping, one can treat impurity and interaction potential as the perturbation to $\hat{H}_0$. Here $H_0$ is the Dirac Hamiltonian coupled to $U(1)$ gauge field,
   %Upon considering the Hamilonian by $\hat{H}=\hat{H}_0+\hat{V}_{\text{imp}}+\hat{V}_{\text{int}}$,  
  \begin{eqnarray} 
  \hat{H}_0  =v_F\int d^2\mathbf{r} \hat\Psi^\dagger(\mathbf{r})   [  \hat \Sigma_\alpha (-i \partial^\alpha+eA^\alpha)  ] \hat\Psi(\mathbf{r}).\end{eqnarray}  
  Here $\alpha $ is summed in $x$ and $y$, $v_F$ is the Fermi velocity, $\hat\Psi=(\hat\psi_{AK},\hat\psi_{BK},\hat\psi_{BK'},\hat\psi_{AK'})$ is the 4-component fermion operator and $\hat\Sigma_{x,y}= \hat{\tau}_z \otimes  \hat{\sigma}_{x,y}$, where $\hat\tau_z$ is the third Pauli matrix acting in $K,K'$ space and $\hat\sigma_{x,y}$ are Pauli matrices acting in the space of $A,B$ sublattices. The gauge field is adopted by $\mathbf{A}=(-eB y,0)$.

 \begin{figure}
	\centering
	\includegraphics[scale=0.5]{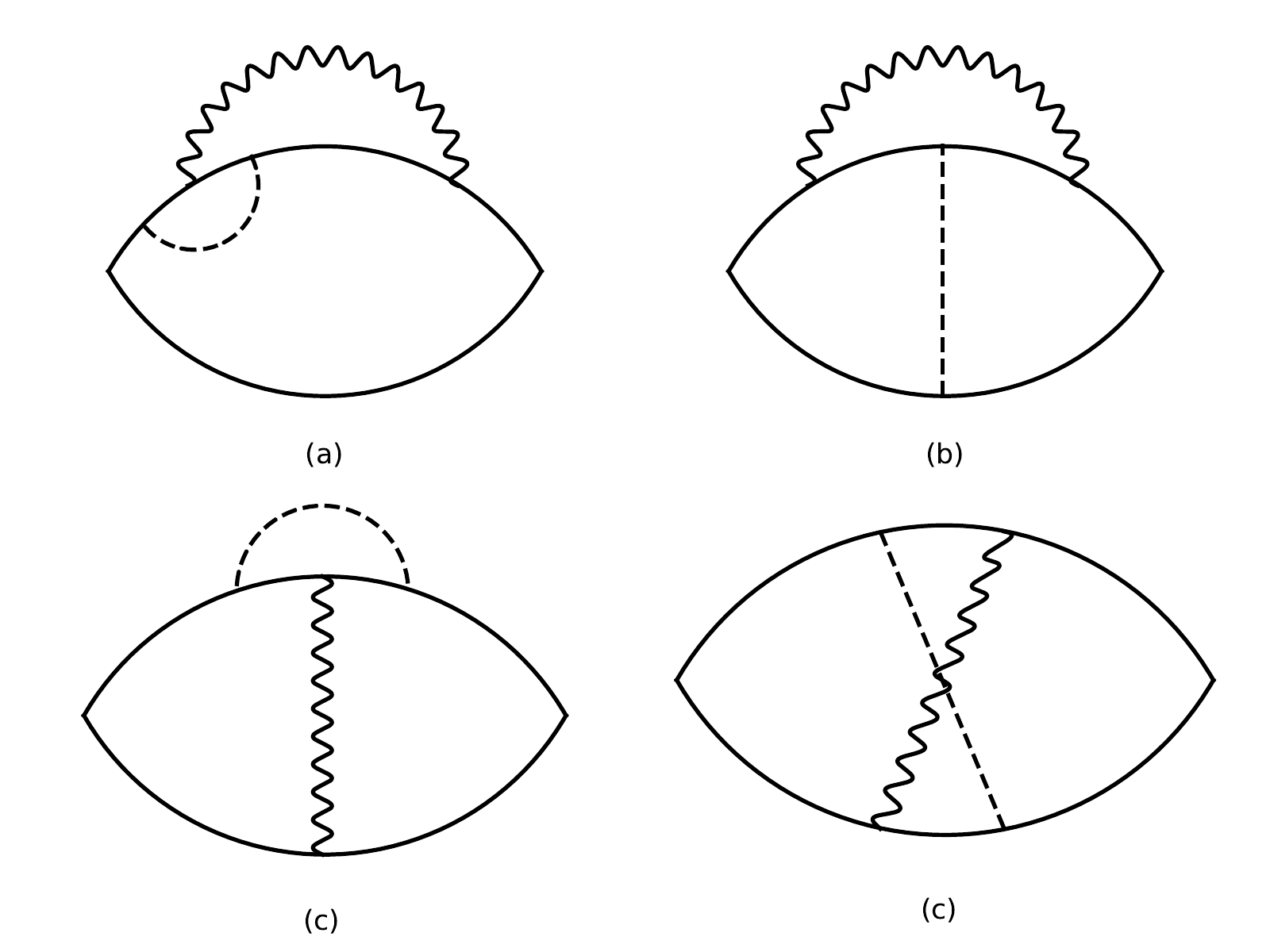} 
	\caption{Feynman diagrams giving leading field-dependent corrections to the longitudinal static conductivity. Solid lines represent the Feynman propagators. Dashed lines are the static impurities, while the wavy lines represent the electron-electron interactions.  }
	\label{diagram}
\end{figure} 
  
Now, we consider the Gaussian-correlated potential. Meanwhile, the symmetry-allowed disorder potential is described by five parameters\cite{prl06Aleiner,prl06Altland,prb06Ostrovsky}, namely,
  \begin{eqnarray}
  \label{correlation_impurity}
  \langle \hat V_{\mathbf{r}} \otimes \hat V_{\mathbf{r}'}  \rangle = \delta_{\mathbf{r},\mathbf{r'}} \Big[
\gamma_0 \hat I\otimes \hat I+ g^m_i \hat Q_m^i\otimes \hat Q_m^i \Big]
  \end{eqnarray}
Here  $\hat V_{\mathbf{r}} $ is the impurity potential and the bracket $\langle  ... \rangle$ is the average over impurity distributions. $\hat{I}$ is the identity matrix. Here $\hat Q_m^i=\hat \Sigma_m \hat \Lambda_i$ and $\hat \Sigma_z=\hat \tau_0\otimes \hat \sigma_z  , 
\hat \Lambda_x=\hat \tau_x \otimes \hat \sigma_z,  \hat \Lambda_y=\hat \tau_y \otimes \hat \sigma_z,  \hat \Lambda_z=\hat \tau_z \otimes \hat \sigma_0
$. We adpot the notation from Ref.~\onlinecite{prl06Aleiner}, $g_z^{z}=\gamma_z$, $g_z^{x,y}=\gamma_\perp$, $g^z_{x,y}=\beta_z$ and $g^{x,y}_{x,y}=\beta_\perp$. { Effectively, $\gamma_0$ represents the square of the strength of static electric potential averaged over the A/B sublattice. Parameters $\beta_z$ and $\beta_\perp$ introduce the intervalley scatterings while $\gamma_\perp$ introduces the hopping between A and B sublattices. The parameter $\gamma_z$ creates a chemical potential difference between the A/B sublattices. To clarify terminology, we refer to the impurity potential from $\gamma_0$ term as the scalar potential while all other terms in the potential as non-diagonal.}  
  
To illustrate coherent scattering quantitatively, we use the semiclassical expression of Dirac propagators in the real space\cite{wang2021}, $
  \langle G(\mathbf{r},\omega) \rangle\sim  e^{i\text{sgn}(\omega) \Phi_0(r)-r/(2\tau v_F) }M(r,\text{sgn}(\omega))/k_Fr
$. Here $\Phi_0(r)$ is the phase including both $k_Fr$ and the magnetic phase\cite{prl07Sedrakyan,random}.  The form of matrix $M$ shows that chiral-symmetry is broken in each valley but it is preserved in the Brillouin zone\cite{wang2021}. The field-dependent part in $M$ reads, $M-M_0\simeq  -{\varphi^2(r)}  \hat{I} /2
   -i\text{sgn}(\omega)\varphi(r)\hat{\Sigma}_z$. Here $M_0$ is the value of matrix $M$ in the absence of field and $\hat{I}$ is the identity matrix. 
  
  Applying perturbations, one finds that a series of Feynman diagrams led to dominant contributions to the conductivity. Up to the lowest orders of the impurity potential and interactions, we find that the diagrams in Fig.~\ref{diagram} give the leading field-dependent corrections to the longitudinal and static conductivity, $\delta \sigma_{xx}$. Namely, these are diagrams that contain vertex corrections\cite{wang2021}, while others in the same order of perturbation theory are subleading. 
  
  The exact expression corresponding diagrams in Fig.~\ref{diagram} can be simplified. In the leading in $(T\tau)^{-1}$ order, and upon neglecting highly-oscillatory $\sim \exp i2k_Fr$ terms, one arrives at a short expression for the conductivity correction\cite{appendix},
 \begin{eqnarray} 
\label{m8}
\!\!\delta \sigma_{xx} \simeq  - \lambda_0 \frac{e^2\tau }{\pi^2 \alpha_{\text{tr}}  }  \int_0^{E_F} \!\!\! d\Omega \frac{d}{d\Omega}\Big(\Omega \coth \frac{\Omega}{2T} \Big) \text{Im} I(\Omega). 
\end{eqnarray}
Here the function $I(\Omega)$ is expressed by the integral, $ I(\Omega)=\int y^{-1}{dy} 
\Big(
2p' \varphi^2(y)   +t'
\Big) e^{2i(\Omega+i\tau) y /v_F}$, $\lambda_0=k_FU_0/2\pi v_F$ is the dimensionless interaction constant at zero momentum.  Parameters $p'$ and $t'$ are defined by $p'=\gamma_0-\beta_z-\gamma_z$ and $t'=2\gamma_z+2\beta_z+\beta_\perp+\gamma_\perp$. The expression of $I(\Omega)$ originates from the coherent scatterings in Fig.~\ref{scattering}.   Notice the constant $t'$ does not contain $\gamma_0$, while $p'$, representing the enhancement of backscattering, depends on $\gamma_0$. 
For short-ranged and weak scatterers, the parameter  $\alpha_{\text{tr}}$ is found to be $\alpha_{\text{tr}}=\gamma_0 + {4\beta_\perp} + { 2\gamma_\perp} + { 2\beta_z} + { \gamma_z}$, determining the Drude conducitivty in graphene\cite{prb06Ostrovsky}.     

The integration in the expression of $\text{Im} I(\Omega)$ can be analytically performed, giving
\begin{eqnarray} 
\label{m9}
\text{Im}I(\Omega)=\frac{\pi t}{2}+ p \frac{(\omega_0\tau)^2}{4}  \frac{ \Omega \tau}{(1+ \Omega^2\tau^2)^2}. 
  \end{eqnarray}
The zero-field part shares the same integration as in Ref.~\onlinecite{01prbzala}. The linear $T$-dependence in $\delta \sigma_{xx}$ at the zero-field is obtained from the property, $\lim_{\Omega\rightarrow 0} \Omega \coth \Omega/2T\simeq 2T$. The sensitivity to the disorder potential in the zero-field limit, namely the sensitivity to parameter $t$, agrees well with the result in Ref.~\onlinecite{prl06Cheianov}.

The field-dependent correction in Eq.~\ref{m8} mainly originates from the region  $0<\Omega<2T$. In this region, one can linearize $\frac{d}{d\Omega}\Big(\Omega \coth \frac{\Omega}{2T} \Big)\simeq \Omega/3T$. Notice that the charateristic scale for $\Omega$ in Eq.~\ref{m9} is $\sim \tau^{-1}$. The integral over $\Omega$ does not introduce extra temperature dependence. Thus the field-dependnet correction $\delta \sigma_{xx}$ is $\sim \omega_0^2T^{-1}$. Tracing the integral rigorously, one can obtain Eq.~\ref{int2}, where we define $t=t'/\alpha_{\text{tr}}$ and $p=p'/\alpha_{\text{tr}}$. Inverting the magnetoconductivity tensor gives us Eq.~\ref{resistivity}, which is the main result of the present work. The result is specific for Dirac electron and valid in a large parameter space when $\omega_0<T<E_F$. { In Ref.~\onlinecite{appendix}, we provide the comparison between the mechanism reported here and the hydrodynamics description\cite{PhysRevB.93.195430,PhysRevB.98.125111,PhysRevB.104.075443,prl16Alekseev,PhysRevB.97.085109,PhysRevB.103.075303,prl17Scaffidi,PhysRevB.103.125106,Lucas_2018}, which is the recent focus of studies on the transport of graphene. We show in Ref.~\onlinecite{appendix}, that for the doped and ultra-clean sample, the reported ballistic magnetotransport mechanism is the dominating effect when the temperature, $T$, within the logarithmic accuracy, lies in a parametrically large interval $\left(\frac{k_B}{\hbar \tau}\right) \lesssim T\lesssim \left(\frac{k_B}{\hbar \tau}\right)\left[\frac{\hbar E_F\tau}{k_B \ln(\hbar E_F\tau/k_B)} \right]^{1/2}$ (from now on we restore $k_B /\hbar$ prefactor in the expression for $T$ ).}

%Temperature dependence in the static conducitivity of graphene is known to be linear like and sensitive to disorder-potential, because of the Berry phase $\pi$ and chiral symmetry of Dirac eletron. The application of a weak magnetic field breaks the chiral symmetry and introduce a novel correction to the conductivity, $\delta \sigma_{xx}\propto \omega_0^2/T$. The result is specific for Dirac electron and valid in a large parameter space when $\omega_0<T<E_F$.

The present technological capabilities do not allow one to engineer the graphene samples with a given impurity type to the best of our knowledge. Therefore, the present theory allows extracting the information about the impurity in the sample from the magnetotransport measurement. Namely, upon fitting the temperature dependence of observed magnetoresistance with Eq.~\ref{resistivity}, one can extract the ratio of $p/t$. This helps to understand if the impurity in the given sample is mostly scalar type $(p/t\gg 1)$ or mostly non-diagonal $(p/t\lesssim 1)$.
 
{\em Implications for the experiments.}  The new magnetoresistance behavior can be observed in experiments, provided with two conditions on disorder: (1). the disorder in the sample should ensure the inequality, $p/t\gg1$. (2). The sample should be clean enough so that the ballistic transport can be observed.

To ensure $p/t\gg 1$, the type of disorder in a sample needs to be primarily scalar-like. Namely, only a small portion of disorder potentials create intra-valley scatterings $A\leftrightarrows B$, intervalley scatterings $K\leftrightarrows K'$, and different on-site chemical potentials on sub-lattices. %In the Ref, the scalar-like potential in Eq.~ can be realized by the case randomly distributed in $A$ and $B$ sublattice

The ballistic transport sets a lower bound for temperature, $ T  >T_0\equiv k_B /\hbar\tau$. Meanwhile, the temperature should be low enough so that the thermal effects and phonon effects do not defeat the quantum effects of electrons. Thus sample should be clean enough for $T_0$ to represent a low temperature.
In the previous {\em magnetotransport} experiments, samples under consideration were not clean enough for the ballistic transport to be observed. For example, 
in Refs.~\onlinecite{prb11Jouault} and \onlinecite{prl12Jobster}, the mobility of sample is $\mu \sim 2\times 10^3 \text{ cm/Vs}$ and the transport time is $\tau\sim 100 \text{ fs}$. The temperature  $T_0$ is $T_0\sim 500 K$.  This is a high temperature where thermal, and phonon effects\cite{prb08hwang,prb08Fratini,iop12Denis,prl10Kim,science10Jae,prb21Sedrakyan} are strong and dominating. Nowadays, a clean sample with highly mobile electrons can be fabricated. According to Ref.~\onlinecite{sa15Luca}, the method of chemical vapor deposition on reusable copper can be used to fabricate the graphene device with a high mobility, $\mu \sim 3.5 \times 10^{5} \text{ cm/Vs}$. The subsequent work\cite{nl16Banszerus} shows that the electron mobility can be enhanced to be $\mu \sim   3\times 10^{6} \text{ cm/Vs}$ together with the observation of ballistic transport at $1.7K$. These recent techniques may allow one to study the magnetoresistance of the doped graphene in the ballistic regime\cite{08mpbandrei,Du2008}. In this regard, the predicted phenomenon in this letter can be observed.

{\em Acknowledgments.} We are grateful to M. E. Raikh for valuable discussions. The research was supported by startup funds from the University of Massachusetts, Amherst. 

%Importantly, when the disorder potential is scalar-like, then $t=0$ and $p\neq 0$. The linear temperature dependence vanishes and the dominating behavior\cite{footnote2} in the interaction correction is reciprocal in $T$.

%Magnetoresistance can be obtained from Eq.~\ref{12} via the relation,$\delta \rho_{xx}\simeq \rho_0^2(\omega_0^2 \tau^2 -1)\delta \sigma_{xx}.$ The field dependent term in
%$\delta\rho_{xx}$ now contains $\sim t\omega^2_0 \tau^2 T+p \omega_0^2 (48 T)^{-1}.$ If the disorder potential obeys $p/t \gg (T\tau)^{2}$ (perturbatively around scalar potential), one can clearly observe that $\omega^2_0T^{-1}$ is the dominating behavior in magnetoresistance.

%In fact, if we have $p/t\gg 1$, we can observe an interesting T-dependent performance in $\delta \rho_{xx}$. Namely, (i) when $\tau^{-1}< T<\sqrt{p/t} \tau^{-1}$, $\delta \rho_{xx}$ is inverse function of $T$. (ii) When $\sqrt{p/t} \tau^{-1}<T< E_F$, the temperature dependence in $\delta \rho_{xx}$ is linear-like. (iii) when $T\tau$ is around $\sqrt{p/t}$, there is a minimal point for $\delta \rho_{xx}$.

	\bibliography{cond}

\begin{thebibliography}{66}
\bibitem [{foo()}]{footnote3}%
  \BibitemOpen
  \href@noop {} {\bibinfo  {journal} {One may notice that
  $\omega^2_0(T\tau)^{-1}$ dependence can also originate from the interaction
  correction to Hall coefficient. See PhysRevB.64.201201. Note that the
  disorder parameter before this correction is ${t}$ and thus the correction is
  always sub-leading to $\omega^2_0(T\tau)$}\ }\BibitemShut {NoStop}%
\end{thebibliography}

    \pagebreak
    \widetext
    \begin{center}
    	\textbf{\large Supplemental Material:   Ballistic magnetotransport in graphene}
    \end{center}
    %%%%%%%%%% Merge with supplemental materials %%%%%%%%%%
    %%%%%%%%%% Prefix a "S" to all equations, figures, tables and reset the counter %%%%%%%%%%
    \setcounter{equation}{0}
    \setcounter{figure}{0}
    \setcounter{table}{0}
    \setcounter{page}{1}
    \makeatletter
    \renewcommand{\theequation}{S\arabic{equation}}
    \renewcommand{\thefigure}{S\arabic{figure}}
    \renewcommand{\bibnumfmt}[1]{[S#1]}
    \renewcommand{\citenumfont}[1]{S#1}

    	\title{
    	Supplementary Material:
    	Ballistic magnetotransport in graphene}
    \author{Ke Wang}
    \email{kewang@umass.edu}
    \affiliation{%
    	Department of Physics, University of Massachusetts, Amherst, MA 01003, USA}
    %\author{M. E. Raikh}
    %\affiliation{Department of Physics and Astronomy, University of Utah, Salt Lake City, UT 84112, USA}
    \author{T. A. Sedrakyan}%
    \email{tsedrakyan@umass.edu}
    \affiliation{%
    	Department of Physics, University of Massachusetts, Amherst, MA 01003, USA}
    
    \maketitle
    
    \section{Basic formulas for static conductivity}
    In this section, we present the definitions and the main formulas for static conductivity. The Dirac electron in graphene, coupled to $U(1)$ gauge field, is described by the Hamiltonian,
    \begin{eqnarray} \hat{H}_0  =v_F\int d^2\mathbf{r} \hat\Psi^\dagger(\mathbf{r})   [   \Sigma_\alpha (-i \partial^\alpha+eA^\alpha)  ] \hat\Psi(\mathbf{r}).\end{eqnarray}  
    Here a summation is assumed over the repeating index, $\alpha $, with $\alpha=x,y$, $v_F$ is the Fermi velocity, $\hat\Psi=(\hat\psi_{AK},\hat\psi_{BK},\hat\psi_{BK'},\hat\psi_{AK'})$ is the 4-component fermion operator. The four-dimensional matrix $\Sigma_{x,y}= {\tau}_z \otimes  {\sigma}_{x,y}$, where $\tau_z$ is the third Pauli matrix acting on $K,K'$ space and $\sigma_{x,y}$ are Pauli matrices acting on the space of $A,B$ sublattices. Then the current operator for Dirac electrons is given by $ ev_F\hat\Psi^\dagger(\mathbf{r})\Sigma_\alpha \hat\Psi(\mathbf{r})$.
    
    Now consider the system with the disorder potential  $V_{\text{imp}}(\mathbf{r})$ and the interaction potential $U(\mathbf{r})$. We assume the correlation of the disorder potential is point-like. Namely,
    \begin{eqnarray} 
    	\label{correlation}
    	\langle   V_{\text{imp}}(\mathbf{r}) \otimes    V_{\text{imp}}(\mathbf{r'}) \rangle_I =\delta_{\mathbf{r},\mathbf{r'}} \Big[
    	\gamma_0 1_4\otimes 1_4+ \beta_\perp \Sigma_{x,y} \Lambda_{x,y} \otimes \Sigma_{x,y} \Lambda_{x,y}  \nonumber\\
    	+\gamma_\perp \Sigma_{x,y} \Lambda_{z} \otimes \Sigma_{x,y} \Lambda_{z}+\beta_z \Sigma_{z} \Lambda_{x,y}\otimes \Sigma_{z} \Lambda_{x,y}
    	+\gamma_z \Sigma_{z} \Lambda_{z}\otimes \Sigma_{z} \Lambda_{z}\Big] 
    \end{eqnarray}
    The bracket $\langle  ... \rangle_I$   is the average over impurity distributions. %Here $\alpha_0$ describe the strength of disorder potential, $\delta_{\mathbf{r},\mathbf{r'}}$ is the delta function and $I$ is the $4\times 4$ identity matrix. 
    Matrices above are defined by  
    $ \Sigma_z=\tau_0\otimes \sigma_z  , 
    \Lambda_x=\tau_x \otimes \sigma_z,  \Lambda_y=\tau_y \otimes \sigma_z,  \Lambda_z=\tau_z \otimes \sigma_0.
    $. The $\gamma_0$, $\beta_\perp$, $\gamma_\perp,\beta_z $ and $\gamma_z$ describe the strength of disorder potential.
    %The impurity-averaged propagator is almost same as the free one,  \begin{eqnarray} \langle G_{R/A} (\mathbf{k},\omega) \rangle_{\text{imp}} =\frac{1+\hat{k}\cdot\hat\Sigma}{\omega-v_F k\pm\frac{i}{2\tau}},  \end{eqnarray}since the lattice symmetries are restored after the average over impurities is done.
    
    According to the Kubo formula, the static conductivity is estimated by the current-current correlation function, $\sigma_{\alpha,\beta}
    =\lim_{\omega\rightarrow 0}\frac{i}{\omega} \Pi_{\alpha,\beta}(\omega)$. Here the $\Pi_{\alpha,\beta}(\omega)$ is obtained by taking analytic continuation of $\Pi_{\alpha,\beta}(i\Omega_n)$ via $i\Omega_n \rightarrow \omega$, 
    \begin{eqnarray}
    	\Pi_{\alpha,\beta}(i\Omega_n)= \int_0^{1/T} d\tau  \langle T_\tau \hat{j}_\alpha(\tau) \hat{j}_\beta(0)  \rangle e^{i\Omega_n\tau}  
    \end{eqnarray}
    Here $\hat{j}_\alpha(\tau)$, $\alpha=1,2$, is the current operator at imaginary time $\tau$, $\omega_n=2\pi T n$ is the bosonic Matsubara frequency and $i\Omega_n \rightarrow \omega$ represents the analytic continuation.  
    
    We treat the electron-electron interaction as the perturbation to $H_0$. The current-current correlation function can be generally expressed by
    \begin{eqnarray} 
    	\Pi_{\alpha,\beta}(i\Omega_n)=-T\sum_{i\omega_m} J_\alpha G(i\omega_m)J_\beta G(i\omega_m-i\Omega_n) 
    	-T\sum_{i\omega_m} J_\alpha G(i\omega_m)\Gamma_\beta(i\omega,i\omega-i\Omega_n) G(i\omega_m-i\Omega_n)  \nonumber\\
    \end{eqnarray}
    where $\omega_m=(2\pi m+1) T $ is the fermionic Matsubara frequency and $\Gamma$ is the vertex correction. The current operator of Dirac electrons $J_\alpha$ above reads $J_\alpha=ev_F \Sigma_\alpha$. Here $G$ is the exact green function for the interacting system. The first term in the RHS is of the self-energy type and the second term contains the vertex correction. Below we refer to the self-energy/vertex-type contributions to $\sigma_{\alpha,\beta}$ as to $\sigma^{S/V}_{\alpha,\beta}$.

    \subsection{Self energy correction}
    The standard analytic continuation procedure transform the summation over Matsubara frequency in $\sigma^{S}_{\alpha,\beta}$ into a single variable integral,
    \begin{eqnarray} 
    	\label{5}
    	\sigma^{S}_{\alpha,\beta}&=&\frac{1}{4\pi}\text{Re}\int_{-\infty}^{+\infty} d\epsilon  \partial_\epsilon\left[ \tanh\frac{\epsilon}{2T} \right] 
    	\Big[  J_\alpha G_R(\epsilon)J_\beta G_A(\epsilon) - J_\alpha G_R(\epsilon)J_\beta  G_R(\epsilon)\Big]   \label{13}.
    \end{eqnarray}
    Here $G_{R/A}$ is the retarded/advanced Green's function. Consider the first order perturbation over the interactions. There are two types of diagrams, Hartree and Fock. The diagrams $a$ and $b$ in Fig.~3 of maintext are of the Fock type. Here, we only focus on the calculation of Fock self energy, $\Sigma_F(i\omega_m)=T \sum_{i\nu_l}  G_0(i\omega_m-i\nu_l) V(i\nu_l)$. Here $\nu_l$ is the bosonic Matsubara frequency, $G_0$ is the Green's function for non-interacting Hamiltonian $H_0$ and $V(i\nu_l)$ is the interaction potential in the frequency space (up to first order in perturbation, $V(i\nu_l)$ is simply constant $V$, not depending on frequencies). Inserting the self-energy $\Sigma_F$ into Eq.~\ref{5} and performing analytic continuation, one obtains
    \begin{eqnarray}
    	\label{6}
    	\sigma^{S}_{\alpha,\beta}\simeq&&  \frac{V}{4\pi^2}\;\; \text{Im}\int d\Omega \frac{d}{d\Omega}\Big(\Omega \coth \frac{\Omega}{2T} \Big)   \Big[  J_\alpha G_R(\epsilon) V(\Omega) 
    	G_A(\epsilon-\Omega) G_R(\epsilon)J_\beta G_A(\epsilon)   \\
    	&& - J_\alpha G_R(\epsilon)  V(\Omega) 
    	G_R(\epsilon-\Omega) G_R(\epsilon)J_\beta G_A(\epsilon)    - J_\alpha G_R(\epsilon)V(\Omega)  G_A(\epsilon-\Omega) G_R(\epsilon)J_\beta  G_R(\epsilon)\Big].\nonumber
    \end{eqnarray}

    \subsection{Vertex correction}
    Similarly, one could perform the analytic continuation in the vertex correction and find
    \begin{eqnarray} 
    	\label{v}
    	\sigma^{V}_{\alpha,\beta}= &&\frac{T}{4\pi  } \text{Re} \int d\epsilon \Big[\partial_\epsilon\tanh\frac{\epsilon}{2T} \Big]  J_\alpha   
    	\Big[ G_R(\epsilon) \Gamma_\beta(\epsilon+i\delta,\epsilon-i\delta) G_A(\epsilon  )
    	- G_R( \epsilon)\Gamma_\beta(\epsilon +i\delta,\epsilon+i\delta)G_R(\epsilon ) \Big], \nonumber\\
    \end{eqnarray} 
    where $\delta$ is an arbitrary small positive number. Diagrams $c$ and $d$ in Fig.~3 of the maintext are of the vertex type. The first order vertex correction reads  $
    \Gamma_\beta(i\omega_m ,i\omega_m-i\nu_n)=VT\sum_{i\nu_l} G_0(i\omega_m-i\nu_l)J_\beta G_0(i\omega_m-i\nu_l-i\nu_n)
    $. Performing the analytic continuation for $i\omega$ , $i\nu$ and inserting $\Gamma$ into Eq.~\ref{v}, one finds
    \begin{eqnarray} 
    	\label{8}
    	\sigma^{V}_{\alpha,\beta} \simeq && \frac{1}{8\pi^2  } \;\text{Im}\int d\Omega  \frac{d}{d\Omega}\Big(\Omega \coth \frac{\Omega}{2T} \Big) \times J_\alpha  \Big[   2G_R(\epsilon)V(\Omega)G_A(\epsilon-\Omega) 
    	J_\beta G_A(\epsilon-\Omega)G_A(\epsilon)   \nonumber\\ 
    	&&-  G_R(\epsilon) V(\Omega)      G_A( \epsilon-\Omega)  J_\beta G_{A}(\epsilon-\Omega)     G_R(\epsilon  ) \Big]. 
    \end{eqnarray}

    \section{Calculation of the interaction corrections to static longitudinal conducitivity}
    This section provides a detailed calculation to derive the main temperature dependence in the static longitudinal conductivity, $\delta \sigma$.
    The Fock-type diagrams, including impurity scatterings, are shown in Fig.~3 of maintext. Taking all four diagrams into consideration, we find that the leading correction to $\sigma$ in the ballistic regime ($T\tau \gg 1$) is given by
    \begin{eqnarray} 
    	\label{9}
    	\delta \sigma\simeq&&  \frac{V}{2\pi^2} \;\text{Im} \int d\Omega  \frac{d}{d\Omega}\Big(\Omega \coth \frac{\Omega}{2T} \Big)  \nonumber \text{tr}  \Big[  J_\alpha G_R(\epsilon)  {V}_{\text{imp}} G_R(\epsilon)  V(\Omega)
    	G_A(\epsilon-\Omega)   {V}_{\text{imp}}G_A(\epsilon-\Omega) G_R(\epsilon)J_\beta  G_A(\epsilon) 
    	\Big]. \nonumber\\
    \end{eqnarray}
    Here $V_{\text{imp}}$ represents the impurity potential. Thus the main task is to evaluate the Eq.~\ref{9}. Here we adopt the semiclassical limit to estimate Eq.~\ref{9}.
    In the semiclassical limit, the interaction potential acts as a touching potential. The real space Green's function averaged over impurities, reads
    \begin{eqnarray}
    	\label{green}
    	\langle G(\mathbf{r},\omega) \rangle_I=\frac{k_F}{2v_F} \sqrt{\frac{1}{2k_Fr}} e^{i\text{sgn}(\omega) \Phi_0(r)-r/(2\tau v_F) }   M,
    \end{eqnarray}
    where the phase $\Phi_0(r)=k_Fr+{ \omega r}/v_F + \pi/4  -r^3/(24k_Fl^4)$ and the matrix $ M$ is given by
    \begin{eqnarray}
    	\label{3}
    	M(\mathbf{r},\text{sgn}(\omega))\simeq  && \left(\text{sgn}(\omega)+i (2k_Fr)^{-1}\right) \hat{r}\cdot \mathbf{\Sigma}+\hat{I}-i\text{sgn}(\omega)\varphi(r)\hat{\Sigma}_z
    	-\frac{\varphi(r)^2}{2}\hat{I},
    \end{eqnarray}
    where $ {\boldsymbol \Sigma}=(\Sigma_x, \Sigma_y)$ and $\hat{I}$ is the identity matrix. Here $\varphi(r)=\omega_0 r/(2v_F)$ is the half of the angle corresponding to the arc of the Larmour circle with length $r$. The angle $\varphi(r)$ represents the extent of the chiral symmetry breaking, since $\Sigma_z$ anti-commutes with $\Sigma_{x,y}$.
    
    To perform real space integration in the expression for the conductivity, one may make use of the following identity 
    \begin{eqnarray} 
    	\label{12}
    	\int d^2x G_R(\mathbf{x}',\mathbf{x};\epsilon) \hat{\Sigma}_\alpha G_{A}(\mathbf{x},\mathbf{y}; \epsilon)=-\frac{i\tau}{k_F}  
    	\frac{\partial }{\partial {x_\alpha'}}  \Big[G_A-G_R\Big] (\mathbf{x}',\mathbf{y};\epsilon),
    \end{eqnarray}
    where $\alpha=1,2$. This  identity helps us in Eq.~\ref{9} bringing the convolution of two propagators into one around the vertex. Applying Eq.~\ref{12} twice (once for $J_\alpha$ and once for $J_\beta$), one arrives at
    \begin{eqnarray} 
    	\delta \sigma \simeq && - \frac{e^2  \tau^2k_F^2}{64\pi^2 v^2_F y^2}  \frac{1}{2\pi^2} \;\;\text{Im}\int d\Omega \frac{d}{d\Omega}\Big(\Omega \coth \frac{\Omega}{2T} \Big) \;\;\text{tr}\int d^2y    \hat{u}   M(-\mathbf{y};+)M(y;-) \hat{u}  M(-\mathbf{y};-)M(\mathbf{y};+)    
    	e^{2i(\Omega+i/\tau) y /v_F}. \nonumber
    \end{eqnarray} 
    Here $\hat{u}$ is a $4\times 4$ matrix and $\hat{u}\otimes \hat{u}$ inherits the matrix structure from Eq.~\ref{correlation}. Namely, it corresponds to the part contained in the square bracket of Eq.~\ref{correlation}. To further evaluate $\delta \sigma$, one needs to perform angular integration yielding
    \begin{eqnarray} 
    	&&\text{tr} \int d\theta \hat{u}   M(-\mathbf{y};+)M(\mathbf{y};-) \hat{u}  M(-\mathbf{y};-)M(\mathbf{y};+)=32\pi\Big(
    	2p \sin^2\varphi(y )+t
    	\Big).
    \end{eqnarray}
    Here $\theta$ is the angular coordinate of $\mathbf{y}$. Parameters are defined as $p'=\gamma_0-\beta_z-\gamma_z$ and $t'=2\gamma_z+2\beta_z+\beta_\perp+\gamma_\perp$. Linearly expanding $\sin \varphi$, one can write $\delta \sigma$ as the following single variable integral
    \begin{eqnarray} 
    	\delta \sigma \simeq && -  \frac{e^2\tau^2k_F^2}{4\pi^3  v_F^2  }  \int d\Omega \frac{d}{d\Omega}\Big(\Omega \coth \frac{\Omega}{2T} \Big)\text{Im} \int \frac{dy}{y} 
    	\Big(
    	p' \frac{y^2}{2k_F^2l^4}  +t'
    	\Big)e^{2i(\Omega+i\tau) y /v_F}. 
    \end{eqnarray} 
    Now we handle the integral over $y$ firstly and define
    \begin{eqnarray} 
    	\label{25}
    	I(B)=\text{Im} \int \frac{dy}{y} 
    	\Big(
    	p' \frac{y^2}{2k_F^2l^4}  +t'
    	\Big)e^{2i(\Omega+i\tau) y /v_F}.      
    \end{eqnarray}
    The zero field result is simply given by
    \begin{eqnarray} 
    	\label{16}
    	I(0)\simeq 	t' \int_{1/k_F}^{\infty} d  y     
    	\frac{1 }{  y } \sin \frac{2\Omega y}{v_F} = \frac{\pi t'}{2}  +O(\Omega/E_F).
    \end{eqnarray}
    Note that the scalar part $\alpha_0$ of impurity potential does not contribute to $t$, i.e., the zero field conductivity. The field-dependent contribution reads
    \begin{eqnarray} 
    	I(B)-I(0)=    \frac{p'}{2k_F^2l^4}  \int_{1/k_F}^{v_F/\omega_0}    
    	y \sin \frac{2\Omega y}{v_F} e^{-\frac{2y}{v_F\tau} }d  y .
    	\label{magnetic}
    \end{eqnarray}  
    We limit our attention to the limit $\omega_0\tau\ll 1$ so that $\exp (- (\omega_0\tau)^{-1})\simeq 0$. This ensures the convergence of the integral. Also $E_F\tau \gg 1$  is assumed. Then we define $x=2y/v_F\tau$ and rewrite the integral as
    \begin{eqnarray} 
    	I(B)-I(0)\simeq  \frac{p'}{8k_F^2l^4} (v_F\tau)^2 \int^{+\infty}_{0}    
    	x \sin\Big( \Omega\tau x\Big)  e^{- x }d  x. 
    \end{eqnarray}
    Here the lower cut-off is set as zero and the upper cut-off is set to be infinity, since $1/(E_F \tau)\ll  1$ and $1/(\omega_0 \tau)\gg 1$.
    Performing the integral, one obtains
    \begin{eqnarray} 
    	\label{18}
    	I(B)-I(0)=  p' \frac{(v_F\tau)^2}{8 k_F^2l^4}  \frac{ 2\Omega \tau}{(1+ \Omega^2\tau^2)^2}.
    \end{eqnarray}
    Thus, with the help of Eqs.~\ref{16} and \ref{18}, one can write the conductivity as a single variable integral
    \begin{eqnarray} 
    	\delta \sigma \simeq && -  \frac{e^2\tau^2k_F^2}{4\pi^3  v_F^2  }  \int d\Omega \frac{d}{d\Omega}\Big(\Omega \coth \frac{\Omega}{2T} \Big) \Big[
    	\frac{\pi t'}{2}+ p' \frac{(v_F\tau)^2}{4k_F^2l^4}  \frac{ \Omega \tau}{(1+ \Omega^2\tau^2)^2} 
    	\Big].
    \end{eqnarray}
    The remaining task is to simply evaluate the integral above. Now we treat the zero-field and field-dependent contributions separately
    \begin{itemize}
    	\item { Zero-field contribution.} At $B=0$, the integral involved is 
    	\begin{eqnarray} 
    		\int_0^{E_F} d\Omega \frac{d}{d\Omega}\Big(\Omega \coth \frac{\Omega}{2T} \Big)= \Big[
    		E_F\coth(E_F/2T)-2T
    		\Big]. \nonumber
    	\end{eqnarray}
    	Here we set the upper bound for $\Omega$ to be the Fermi energy, $E_F$. This integral leads to the well-known linear temperature dependence in the longtudinal conductivity. Since $E_F/T \gg 1$, the temperature dependence in $\coth(E_F/2T) $function is exponentially weak. The main temperature dependence comes from the latter,the linear one $-2T$.
    	\item {Field-dependent contribution.} The field-dependent contribution is given by the integral 
    	\begin{eqnarray} 
    		\int d\Omega \frac{d}{d\Omega}\Big(\Omega \coth \frac{\Omega}{2T} \Big)  
    		\frac{\Omega \tau}{(1+\Omega^2\tau^2)^2}=2T\int_0^\infty dz \frac{d}{dz}\Big(z \coth z \Big)     \frac{2z\times T\tau }{(1+4z^2 (T\tau)^2)^2} \label{integral_field}.\nonumber\\ 
    	\end{eqnarray} 
    	Here we define the variable $z=\Omega/2T$. Since we consider the ballistic regime, 
    	we only need the asymptotic behavior of the integral at $T\tau \gg 1$. We find that Eq.~\ref{integral_field} has the asymptotic behavior, $\alpha  2T (T\tau)^{-2}$ when $T\tau \gg 1$. Here $\alpha$ is analytically found to be $\alpha= \pi/24$. The functional dependence $\propto (T\tau)^{-2}$ and $\alpha$ are derived below. 
    	
    	For the integrand in the Eq.~\ref{integral_field}, one can separate the integral domain into two parts. They are $(0,\kappa)$ and  $(\kappa,\infty)$ for variable $z$ with $\kappa\sim 1$. We call each region's contribution to the integral as $J_1, J_2$ in a sequence. At the first region, $\partial_z (z/\tanh z) \simeq 2z/3$. Then one defines $x=2z T\tau$ and gets%&\simeq& 2T \int_0^{1/2T\tau} dz  \frac{2z}{3} \frac{2z\times T\tau }{(1+4z^2 (T\tau)^2)^2} =
    	\begin{eqnarray}
    		J_1 = 2T  \frac{(T\tau)^{-2}}{6} \int_0^{+\infty} dx \frac{x^2}{(1+x^2)^2} =\frac{\pi}{24} 2T   (T\tau)^{-2}.  
    	\end{eqnarray}
    	Here we extend the upper bound of integral $2T\tau$ to $\infty$, since $T\tau \gg 1$ and we are looking for the leading order. 
    	In second region, $\partial_z (z/\tanh z) \simeq 1$. One similarly obtains
    	\begin{eqnarray}
    		J_2 \simeq 2T \int_{\kappa}^{\infty} dz  (2zT\tau)^{-3}\propto  (T\tau)^{-3}.
    	\end{eqnarray}
    	From the analysis, one can clearly see that $J_1$ and $J_2$ give the leading contributions and the asymptotic behavior of Eq.~\ref{integral_field} is $\sim (T\tau)^{-2}$.
    \end{itemize}  
    Now we assume the weak and short-ranged scatterer and identity the expression of $\tau$ in the Born-approximation\cite{prb06Ostrovsky},  
    \begin{eqnarray}
    	\frac{1}{\tau}= \frac{k_F}{2  v_F}( {\gamma_0} + {4\beta_\perp} + { 2\gamma_\perp} + { 2\beta_z} + { \gamma_z} ).
    \end{eqnarray}
    Thus $\delta \sigma$ can be simplified to be
    \begin{eqnarray} 
    	\label{conductivity}
    	\delta \sigma\simeq   \lambda\frac{e^2\tau}{\pi  } \Big( \tilde{t} T- \tilde{p} \frac{\omega_0^2}{48T}  \Big).
    \end{eqnarray} 
    Here $\lambda$ is the dimensionless interaction parameter $\lambda=U_0 k_F (2\pi v_F)^{-1}$. Parameters $\tilde{t}$ and $\tilde{p}$ are dimensionless and describe the strength of disorder potential. They are defined by
    \begin{eqnarray} 
    	\tilde{t}=\frac{2\gamma_z+2\beta_z+\beta_\perp+\gamma_\perp}{ {\gamma_0} + {4\beta_\perp} + { 2\gamma_\perp} + { 2\beta_z} + { \gamma_z}},\quad
    	\tilde{p}=\frac{ \gamma_0-\beta_z-\gamma_z }{ {\gamma_0} + {4\beta_\perp} + { 2\gamma_\perp} + { 2\beta_z} + { \gamma_z}}.
    \end{eqnarray}
    Eq.~\ref{conductivity} is the main conclusion in this note. (i) The zero-field contribution linearly depends on the temperature, and this linear dependence is highly sensitive to the nature of the disorder. Once the impurity potential is scalar-like, i.e., only $\alpha_0\neq 0$ while $\gamma_z=\beta_z=\beta_\perp=\gamma_\perp=0$, the linear in temperature term vanishes. (ii) The field-dependent correction is inversely dependent on the temperature. If the impurity potential is scalar-like, the $\omega^2_0/T$ gives the leading interaction correction to the conductivity.
    
    \section{Comparison of the ballistic magnetotransport with the hydrodynamics description}    
    
    Here, we present a comparison between the mechanism in this paper and the hydrodynamics, which has been the main focus of studies recently. We also argue that the mechanism in this paper is the dominating effect when temperature lies in a parametrically large interval.

    It is well established that clean many-body quantum systems may exhibit the hydrodynamic limit\cite{Lucas_2018}. To observe hydrodynamics, the many-body system shall have the condition on scales, $l_{ee}\ll L$. Here $l_{ee}$ is the mean free path due to the interaction while $L$ is the one due to disorder/impurity potential. In other words, the length scale from interactions shall be the smallest.  
    
    We mainly focus on the effects of interactions on magnetotransport of doped graphene. This implies that the Femi energy is the largest energy scale, $\hbar E_F\tau/k_B\gg 1$, so the quasiparticles are long-lived and weakly interacting. This indicates that $l_{ee}$ is very large. Moreover, even in the ultra-clean sample, 
    %$l_{ee}$ can be much larger than the $l_{\text{dis}}$ when the %temperature lies in a parametrically large interval. 
    %Thus the result reported in this letter can be stronger than effects %originating from the hydrodynamics.
    One can show that  $l_{ee}>L$ in a parametrically large temperature interval. Consider a temperature scale corresponding to ballistic transport given in terms of the  impurity scattering time, $\tau$,  
    \begin{eqnarray}
    	T_0= \hbar L/v_Fk_B.
    \end{eqnarray}
    In doped graphene the scattering rate of quasi-particles, $1/\tau_{ee}$, due to the interaction is determined by the Fermi-liquid processes to be \cite{prb82Quinn,prb07das,prl07}
    \begin{eqnarray}
    	\hbar\tau^{-1}_{ee}\sim  \frac{k^2_BT^2}{E_F} \ln \frac{E_F}{T}.
    \end{eqnarray}
    The condition  $l_{ee}\gg L$ is then satisfied at temperstures, $T$, such that
    \begin{eqnarray}
    	\frac{k^2_BT^2}{E_F} \ln \frac{E_F}{k_BT} <k_B T_0.
    	\label{5}
    \end{eqnarray} 
    This condition implies that within the logarithmic accuracy $T\lesssim \left(\frac{k_B}{\hbar \tau}\right)\left[\frac{\hbar E_F\tau}{k_B \ln(\hbar E_F\tau/k_B)} \right]^{1/2}$.	The estimate for the temperature can be obtained from Eq.~(\ref{5}) more accurately. It yields
    \begin{eqnarray}
    	\frac{\alpha^2}{\alpha_0} \ln (1/\alpha) <1
    	,\quad \text{with }  \alpha=\frac{k_B T}{E_F}, \quad \alpha_0=\frac{k_B T_0}{E_F} 
    \end{eqnarray}
    Under this condition, the ballistic magnetotramsport reported in the present letter is dominating over the hydrodynamics mechanism. Note the condition can be applied to any sample, for given disorder-related parameter, $\alpha_0$, of the sample.

    For samples studied in Refs.~\onlinecite{sa15Luca} and \onlinecite{nature19}, one has
    \begin{eqnarray}
    	\label{31}
    	E_F\simeq 0.4 eV,\quad T_0= \hbar L/v_Fk_B\simeq 1.7K
    \end{eqnarray}where $E_F$ is the Fermi energy, $v_F$ is the Fermi velocity and $k_B$ is the Boltzman constant.  Given the data, one can evaluate the corresponding $\alpha_0\simeq 3.6\times 10^{-4}$. Then the condition (\ref{31}) on the temperature  acquires the form	
    \begin{eqnarray}
    	\label{32}
    	f(\alpha)\equiv  \alpha^2\times  \ln  (1/\alpha) <\alpha_0\simeq 3.6\times 10^{-4}.
    \end{eqnarray}
    Fig.~\ref{diagram} depicts the parameter range for temperature under which the condition is satisfied, namely when $\alpha<8.7\times 10^{-3}$. This means for samples under consideration $T<40 K$. Thus the main result of the present letter is dominating for the ultra-clean and doped sample in Eq.~\ref{31} , when the temperature lies in a parametrically large interval, { $1.7 K <T<40 K$}. Here the lower bound is set by the ballistic regime.   
    
    \begin{figure}
    	\centering
    	\includegraphics[scale=0.7]{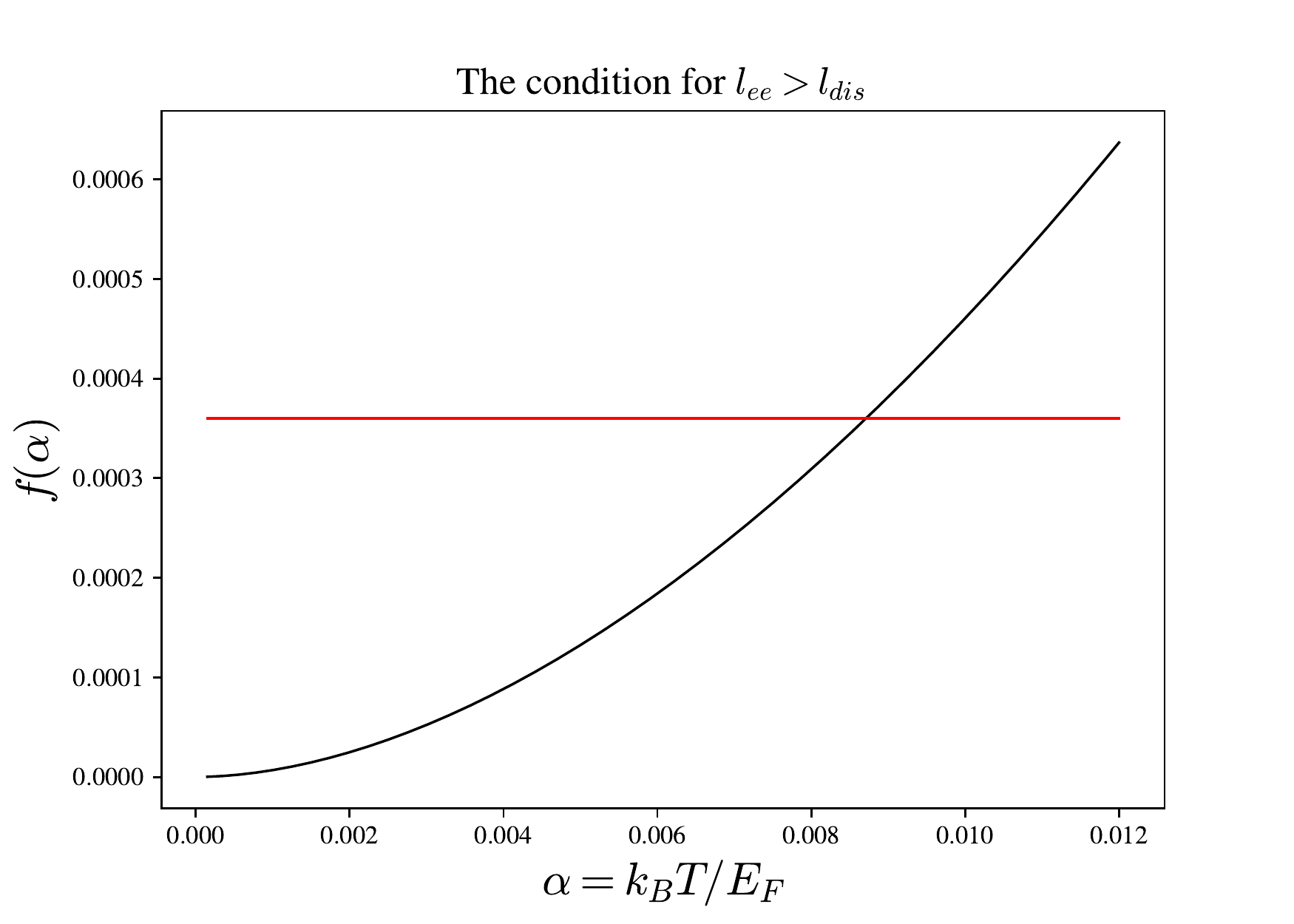} 
    	\caption{ (color online) The plot of $f(\alpha)$ defined in Eq.~\ref{32}. The red line locates the value of $\alpha_0$. The intersection of two curves is around $\alpha\simeq 8.7\times 10^{-3}$. }
    	\label{diagram}
    \end{figure} 
    
\end{document}